\documentclass[%
 aip,
 apl,
 amsmath,amssymb,
 reprint,%
]{revtex4-2}

\usepackage{graphicx}
\usepackage{dcolumn}
\usepackage{bm}

\usepackage[utf8]{inputenc}
\usepackage[T1]{fontenc}
\usepackage{mathptmx}
\usepackage{etoolbox}

\makeatletter
\def\@email#1#2{%
 \endgroup
 \patchcmd{\titleblock@produce}
  {\frontmatter@RRAPformat}
  {\frontmatter@RRAPformat{\produce@RRAP{*#1\href{mailto:#2}{#2}}}\frontmatter@RRAPformat}
  {}{}
}%
\makeatother
\begin{document}


\title{Spin-torque microwave detectors of positive rectangular pulse signals}

\author{V. Prokopenko}
\homepage{v.o.prokopenko@gmail.com}

\author{O. Shtanko}%
\author{I. Sotnyk}
\affiliation{%
Educational and Scientific Institute of High Technologies, Taras Shevchenko National University of Kyiv,
Kyiv 01601, Ukraine
}%

\author{O. Prokopenko}
\homepage{oleksandr.prokopenko@gmail.com}
\affiliation{%
Educational and Scientific Institute of High Technologies, Taras Shevchenko National University of Kyiv, Kyiv 01601, Ukraine
}%
\affiliation{G. V. Kurdyumov Institute for Metal Physics, National Academy of Sciences of Ukraine, Kyiv 03142, Ukraine}

\date{\today}

\begin{abstract}
We analyze the performance of a spin-torque microwave detector (STMD) driven by positive rectangular current pulses $I(t)$ of various amplitudes $I_0$, durations $\tau$, and repetition periods $T$ and reveal two distinct regimes of STMD operation. 
In the first (linear) regime, the time-averaged voltage across the detector, $U_{\rm dc}$, changes linearly with the pulse amplitude $I_0$ and depends on the ratio $\tau/T$: $U_{\rm dc} \sim I_0 (\tau/T)$. 
This regime is observed for a wide range of pulse amplitudes $I_0$ in the case of in-plane (IP) magnetization dynamics and for rather small pulse amplitudes $I_0 \le I_{\rm th}$ in an STMD with out-of-plane (OOP) magnetization dynamics. 
The other (nonlinear) regime is characterized by voltage jumps and drops and is observed only in a structure with OOP magnetization dynamics for input pulses with short repetition periods and large amplitudes $I_0 \ge I_{\rm th}$. 
We believe that the linear regime of STMD operation can be used to unambiguously detect input pulse parameters, which could be important for the development and optimization of spintronic devices capable of detecting and processing non-harmonic (e.g., digital) microwave signals.
\end{abstract}

\maketitle

Nowadays, spin-torque microwave detectors (STMDs) based on magnetic tunnel junctions (MTJs) and utilizing the spin-torque diode effect \cite{Tulapurkar2005Nat} are considered promising nano-scale detectors of microwave signals that can be useful for a wide range of applications in communications, medical imaging, energy harvesting, material characterization, efficient computing, military technology, etc. \cite{Tulapurkar2005Nat,Prokopenko2012JAP,Prokopenko2013Book,Fang2019PRAppl,Tomasello2020PRAppl,
Finocchio2021APL,Artemchuk2021AIPAdv,Mazza2022PRAppl,Sharma2024NatEl,Liu2026NatNano,Kurebayashi2026NatRevPhys}
Their noise-handling properties \cite{Prokopenko2011APL,Prokopenko2013Book} are similar to those of semiconductor detectors of microwave signals (e.g., Schottky diode-based detectors), and their fabrication technology is compatible with semiconductor technology \cite{Liu2026NatNano}. 
Additionally, unlike semiconductor devices, STMDs can be substantially more sensitive to weak signals ($\sim 10$ to $10^3$ times better) in certain operating regimes \cite{Zhang2018APL,Goto2021NatCommun,Zhang2023APL,Thilakaraj2026PhysScr}, paving the way for their extensive use in next-generation communication, medical, and energy harvesting systems.

However, despite many years of rigorous theoretical, numerical, and experimental study of STMD properties, this research has focused only on the case of an input harmonic (sine-wave) current, $I_0 \sin(2\pi f t + \phi)$ 
(see Refs.~\cite{Tulapurkar2005Nat,Prokopenko2012JAP,Prokopenko2013Book,Fang2019PRAppl,Tomasello2020PRAppl,
Finocchio2021APL,Artemchuk2021AIPAdv,Mazza2022PRAppl,Sharma2024NatEl,Liu2026NatNano,Kurebayashi2026NatRevPhys,
Prokopenko2011APL,Zhang2018APL,Goto2021NatCommun,Zhang2023APL,Thilakaraj2026PhysScr}), or a small set of harmonic signals \cite{Berkov2024PRAppl}, where $I_0$, $f$ and $\phi$ are the amplitude, frequency, and initial phase of the input signal current, respectively. 
Conversely, the response of an STMD to a \emph{non-harmonic} signal, particularly a \emph{pulsed} signal current with arbitrary periodic time dependence, has been analyzed only fragmentarily, despite the widespread use of pulsed signals in modern electronics and communication systems. 
Currently, only the cases of several sine-wave signals acting on an STMD \cite{Berkov2024PRAppl} and an oscillator-based spectrum analyzer \cite{Louis2018APL} with a slowly changing input sine-wave signal frequency are considered. 
Therefore, it is critical to understand how prospective spintronic detectors (e.g., STMDs) respond to pulsed driving signals frequently used in modern low-power electronics, including Internet of Things sensors \cite{Jamshed2022SensJ} and spiking neural networks \cite{Yamazaki2022BrainSci}.

\begin{figure}
\includegraphics[width=1.0\columnwidth]{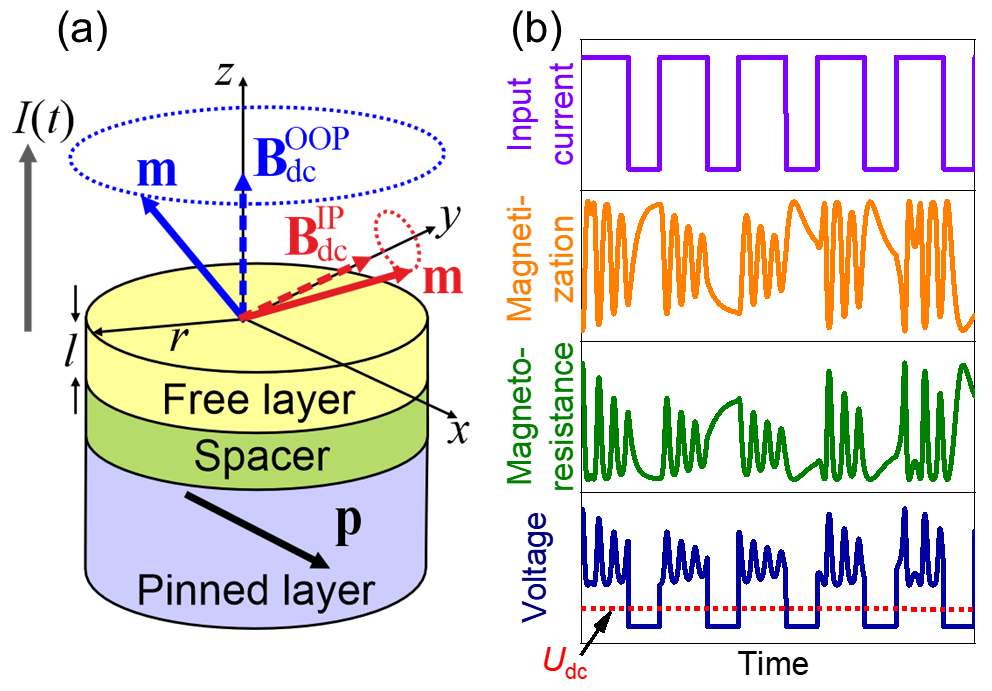}
\caption{
(a) Layout of a circular STMD consisting of free and pinned magnetic layers separated by a dielectric spacer. 
When an input pulse current $I(t)$ is applied to the structure, it excites the magnetization dynamics, ${\bf M}(t)$, described by the normalized magnetization vector ${\bf m} \equiv {\bf m}(t) = {\bf M}(t)/|{\bf M}(t)|$. 
The vector ${\bf m}$ precesses along a particular trajectory, depending on the configuration of the bias dc magnetic field ${\bf B}_{\rm dc}$: an in-plane trajectory (red dotted curve) for an in-plane field ${\bf B}_{\rm dc} = {\bf B}^{\rm IP}_{\rm dc}$ (red dashed arrow) and an out-of-plane trajectory (blue dotted curve) for an out-of-plane field ${\bf B}_{\rm dc} = {\bf B}^{\rm OOP}_{\rm dc}$ (blue dashed arrow).
(b) A set of plots illustrating the principle of the device operation (from top to bottom). 
The input rectangular pulse current $I(t)$, applied to an STMD (top plot), excites the magnetization dynamics in the detector's free layer (second plot), which gives rise to a variation in the device's magnetoresistance $R(t)$ (third plot). 
When the current $I(t)$ flows through the oscillating magnetoresistance $R(t)$, a time-dependent pulsed voltage $U(t)$ is generated across the STMD (bottom plot). 
The average value of this voltage is the output dc voltage of an STMD, $U_{\rm dc}$.
}
\label{f:Model}
\end{figure}

The operation of an STMD of pulsed signals is generally similar to that of an STMD under the action of an input sine-wave signal \cite{Tulapurkar2005Nat,Prokopenko2012JAP,Prokopenko2013Book,Fang2019PRAppl,Tomasello2020PRAppl,
Finocchio2021APL}. 
When an input pulsed signal current $I(t)$ traverses an MTJ, it excites magnetization dynamics in the free magnetic layer (FL) of an STMD structure (Fig.~\ref{f:Model}). 
Since the magnetization of the pinned magnetic layer (PL) remains fixed in the presence of the input signal, the angle between the magnetizations of the FL and PL varies over time, which gives rise to oscillation of the device's tunneling magnetoresistance, $R(t)$ (Fig.~\ref{f:Model}(b)). 
The combination of the current $I(t)$ and magnetoresistance $R(t)$ oscillations yields a time-dependent voltage $U(t)$ generated across the junction that can have a non-zero time-averaged value (Fig.~\ref{f:Model}(b)). 
In analogy with a conventional, harmonic-biased STMD, this value can be treated as the detector output direct current (dc) voltage $U_{\rm dc}$.

In this Letter, we present a numerical and simplified theoretical analysis of the ability of a passive STMD (no bias dc current) to detect input periodic rectangular pulse signals. 
We use STMD models developed in \cite{Prokopenko2012JAP,Prokopenko2013Book,Tomasello2020PRAppl,Artemchuk2021AIPAdv} with an input signal current $I(t)$ in the form of a periodic sequence of positive rectangular pulses. 
Our analysis investigates magnetization dynamics excited by input current pulses $I(t)$ of various amplitudes $I_0$, durations $\tau$, and repetition periods $T$. 
Additionally, we examine how the detector's output dc voltage, $U_{\rm dc}$, can be estimated from the oscillating instantaneous voltage, $U(t)$, generated across the STMD. 
We also test the possibility of identifying the input signal parameters for a known (e.g., experimentally measured) value of $U_{\rm dc}$.

We consider an STMD as a typical, circular, three-layer MTJ structure, in which the free and pinned magnetic layers are separated by a non-magnetic dielectric spacer (Fig.~\ref{f:Model}(a)). 
For simplicity, we assume that the magnetization in the PL is absolutely fixed and directed along the unit vector ${\bf p}=\{1,0,0\} = \hat{{\bf x}}$ (along the $x$-axis in Fig.~\ref{f:Model}(a)). 
Meanwhile, the magnetization dynamics in the FL is governed by the Landau-Lifshitz-Gilbert-Slonczewski equation, which we write using the macrospin approximation \cite{Prokopenko2012JAP,Prokopenko2013Book,Tomasello2020PRAppl,Artemchuk2021AIPAdv}:
\begin{equation}
\label{e:LLGS}
	\frac{d {\bf m}}{d t} 
    = 
    \gamma[{\bf B}_{\rm eff} \times {\bf m}] 
    + 
    \alpha\left[{\bf m} \times \frac{d {\bf m}}{d t}\right] 
    + 
    \sigma I(t)[{\bf m} \times [{\bf m} \times {\bf p}]] 
    \, .
\end{equation}
Here ${\bf m} \equiv {\bf m}(t) = {\bf M}/M_s$ is the normalized magnetization vector associated with the FL magnetization vector ${\bf M} \equiv {\bf M}(t)$, the absolute value of ${\bf M}$, $|{\bf M}| = M_s$, is the saturation magnetization of the FL, $\gamma$ is the absolute value of the gyromagnetic ratio, $\alpha$ is the Gilbert damping parameter, $\sigma = \sigma_\bot / (1 + \eta^2 ({\bf m}\cdot{\bf p}))$ is the spin-transfer-torque coefficient, $\sigma_\bot = (\gamma\hbar/2e) (\eta / M_s V)$, 
$\hbar$ is the reduced Planck constant, $e$ is the elementary electric charge, $\eta$ is the spin polarization efficiency of the current $I(t)$, and $V = \pi r^2 l$ is the volume of the circular FL with radius $r$ and thickness $l$.

The effective magnetic field induction, ${\bf B}_{\rm eff}$, has contributions only from the bias dc magnetic field, ${\bf B}_{\rm dc}$, and the demagnetization field, ${\bf B}_d$, written in the scope of the thin film approximation \cite{MelkovBook}, ${\bf B}_d = - \hat{{\bf z}}\mu_0 M_s ({\bf m}\cdot\hat{{\bf z}})$, where $\hat{{\bf z}}$ is the unit vector of the $z$-axis, and $\mu_0$ is the vacuum permeability.
We consider two cases of the applied bias dc magnetic field: an in-plane field, ${\bf B}_{\rm dc} = {\bf B}^{\rm IP}_{\rm dc} = \hat{{\bf y}}B_{\rm dc}$, which is typical for the in-plane (IP) resonant regime of STMD operation \cite{Tulapurkar2005Nat,Prokopenko2011APL,Prokopenko2013Book}, and an out-of-plane field, ${\bf B}_{\rm dc} = {\bf B}^{\rm OOP}_{\rm dc} = \hat{{\bf z}}B_{\rm dc}$, which is typical for the out-of-plane (OOP) non-resonant regime of STMD operation \cite{Prokopenko2012JAP,Prokopenko2013Book,Fang2019PRAppl,Tomasello2020PRAppl} (here $B_{\rm dc}$ is the field magnitude and $\hat{{\bf y}}$ is the unit vector of the $y$-axis).

The input signal current $I(t) = I_0 i(t)$ in (\ref{e:LLGS}) is a periodic sequence of rectangular pulses with a repetition period $T$, a duration $\tau$ and an amplitude $I_0 > 0$:
\begin{equation}
\label{e:It}
	I(t) 
    =
    I_0 i(t)
    = 
    I_0
    \left\{
    \begin{array}{ll}
          1 \,, &  (n - 1) T \le t \le (n - 1) T + \tau \\
          0 \,, & (n - 1) T + \tau \le t \le n T
    \end{array}
    \right.
    \,,
\end{equation}
where $n$ is the pulse number starting from $1$.

We introduce the time-dependent voltage $U(t)$ generated across an STMD by the input signal current $I(t) = I_0 i(t)$ as 
\begin{equation}
\label{e:Ut}
	U(t) 
    = 
    I(t) R(t)
    =
    I_0 R_\bot
    \frac{i(t)}{1 + \eta^2 ({\bf m}\cdot{\bf p})}
    \, ,
\end{equation}
where $R(t) = R_\bot / (1 + \eta^2 ({\bf m}\cdot{\bf p}))$ is the device magnetoresis\-tance \cite{Prokopenko2013Book}, $R_\bot = {\rm RA}/\pi r^2$ is the MTJ resistance in the perpendicular state when ${\bf m} \bot {\bf p}$, and ${\rm RA}$ is the junction resistance-area product \cite{Prokopenko2013Book} (note that ${\bf m} \cdot {\bf p}$ gives $m_x(t)$).
Then, the output dc voltage of an STMD can be estimated as
\begin{equation}
\label{e:Udc}
	U_{\rm dc} 
    = 
    \frac{1}{T_i} \int_{t_0}^{t_0 + T_i} U(t) dt
    =  
    \frac{I_0 R_\bot}{T_i} \int_{t_0}^{t_0 + T_i} \frac{i(t) dt}{1 + \eta^2 ({\bf m}\cdot{\bf p})}
    \, .
\end{equation}
Here $t_0$ is the time from which the magnetization oscillations can be considered sufficiently stable (we use $t_0 = 10 T$), and $T_i = N T$ is the integration (averaging) time interval, which covers an integer number of pulse repetition periods $N \ge 1$.

We use the following typical parameters of an STMD to quantitatively analyze the device performance \cite{Prokopenko2012JAP,Prokopenko2013Book,Tomasello2020PRAppl,Zhang2023APL,Berkov2024PRAppl}: 
FL radius $r = 50$~nm and FL thickness $l = 1$~nm, spin polarization efficiency of current $\eta = 0.7$, resistance-area product ${\rm RA} = 7.854 \ {\rm \Omega} \cdot {\rm \mu m}^2$, giving $R_\bot = 1 \ {\rm k\Omega}$, Gilbert damping parameter $\alpha = 0.01$, normalized saturation of the FL $\mu_0 M_s = 800$~mT, and the bias dc magnetic field $B_{\rm dc} = 200$~mT. 
The total simulation time is $T_s = 30 T$. 
The parameters of the input pulsed signal vary within the following ranges: pulse amplitude $I_0 \in (0, 0.7]$~mA (the current density varies from $0$ to $8.91\cdot10^6$~A/cm${}^2$), pulse duration $\tau \in [0.1, 0.875]T$ (pulse ratio $T/\tau \in [1.14, 10]$), and pulse repetition period $T \in [0.5, 4]$~ns.

\begin{figure*}
\includegraphics[width=1.0\textwidth]{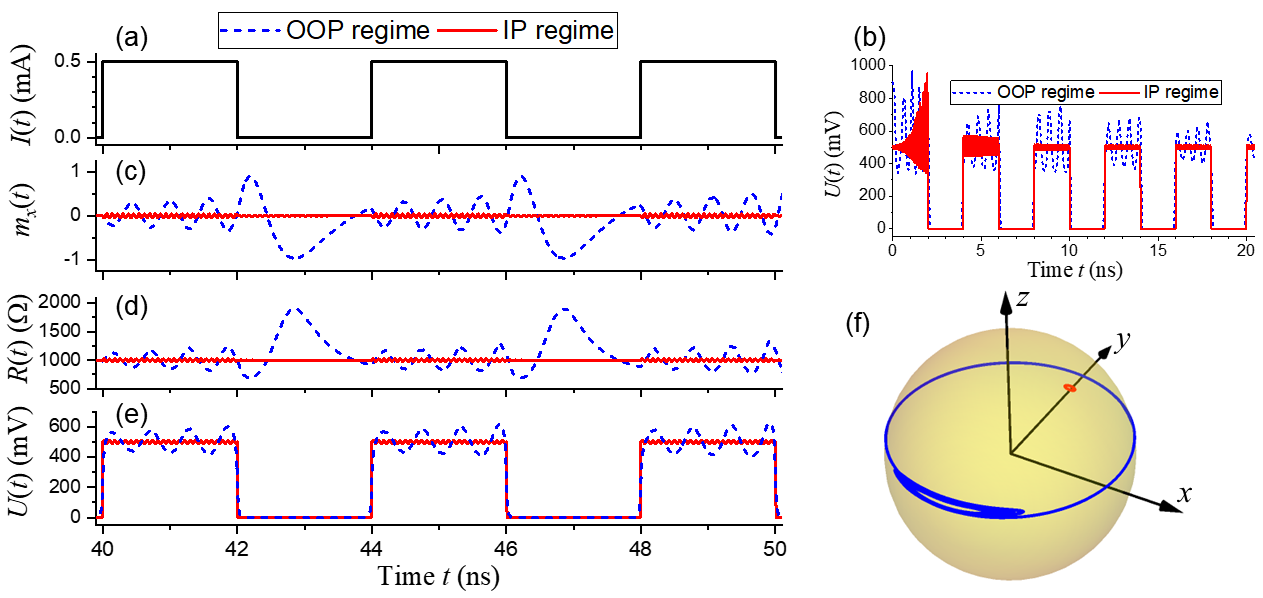}
\caption{
Typical time dependence of (a) the input rectangular pulse current $I(t)$, (c) the $x$-component of dynamic magnetization, $m_x(t)$, (d) the device magnetoresistance $R(t)$, and (b, e) the voltage $U(t)$ generated across the device. 
Solid red lines and dashed blue lines in (b)-(e) correspond to the IP and OOP regimes of STMD operation, respectively. 
Picture (f) shows the trajectory of magnetization motion (red line for an in-plane regime and blue line for an out-of-plane regime) on a surface of unit sphere $|{\bf m}| = 1$. Calculations performed for pulse parameters $I_0 = 0.5$~mA, $\tau = 2$~ns, $T = 4$~ns. All other calculation parameters are indicated in the text.
}
\label{f:ImxRU}
\end{figure*}

Fig.~\ref{f:ImxRU} shows typical simulation results for an input signal $I(t)$ with an amplitude of $I_0 = 0.5$~mA, a pulse duration of $\tau = 2$~ns, and a repetition period of $T = 4$~ns. 
Since the transient dynamics characterize the initial time interval of $0 \le t \le t_0 = 10 T$ (see Fig.~\ref{f:ImxRU}(b) for $0 \le t \le 5 T = 20$~ns), this time interval is excluded from the analysis and data in Fig.~\ref{f:ImxRU}(a, c-e) are presented for for the time period of $t \ge 10 T = 40$~ns only. 
As can be seen, rectangular positive current pulses (Fig.~\ref{f:ImxRU}(a)) excite small quasi-harmonic (red solid line, along the $y$-axis) and large non-harmonic (blue dashed line, along the $z$-axis) oscillations of the FL magnetization, $m_x(t)$, in the IP and OOP regime (see Fig.~\ref{f:ImxRU}(c)), respectively. 
These oscillations lead to small and large time variations of the device magnetoresistance $R(t)$ as shown in Fig.~\ref{f:ImxRU}(d). 
They also result in the generation of a time-dependent, pulse-like voltage, $U(t) = I(t)R(t)$ (see Fig.~\ref{f:ImxRU}(e)). 
The solid red lines and dashed blue lines in Fig.~\ref{f:ImxRU}(b-e) correspond to the IP and OOP regimes, respectively. 
These distinct STMD operation regimes are clearly visible from the magnetization motion trajectories depicted on the surface of the unit sphere, $|{\bf m}| = 1$ (see the red line for the IP regime where ${\bf m}$ oscillates along the $y$-axis, and the blue line for the OOP regime where ${\bf m}$ oscillates along the $z$-axis in Fig.~\ref{f:ImxRU}(f)).

As expected, the generated voltage $U(t)$ has a pulse-like character, mimicking the behavior of the input current $I(t)$, according to the equation $U(t) = I(t) R(t)$.
However, the voltage dynamics differ between the IP and OOP regimes.

In the IP regime, when $m_x(t)$ and $R(t)$ change weakly over time, $U(t)$ can be approximately estimated as $I(t) R_0$, where $R_0$ is the equilibrium (time-averaged) value of $R(t)$. 
Since in the IP regime ${\bf m}$ oscillates along the $y$-axis and ${\bf m} \cdot {\bf p} \approx 0$, $R_0 \approx R_\bot = 1 \ {\rm k\Omega}$.
Therefore, $U(t)$ forms almost ideal rectangular voltage positive pulses with an amplitude of $I_0 R_0 \approx 500$~mV, a pulse duration, and a repetition period equal to those of the input current $I(t)$, i.e., $\tau = 2$~ns and $T = 4$~ns, respectively (see the red solid line in Fig.~\ref{f:ImxRU}(e)).
In this case, according to equation (\ref{e:Udc}), the output dc voltage of an STMD $U_{\rm dc}$ can be estimated quite precisely as 
\begin{equation}
\label{e:UdcApprox}
	U_{\rm dc} 
    \approx 
    q I_0 R_\bot \frac{\tau}{T}
    \, ,
\end{equation}
where $q = 1$ for the considered IP regime.
The obtained equation illustrates that in the IP regime, when the current excites small oscillations of the FL magnetization, the output dc voltage $U_{\rm dc}$ should increase linearly with the input pulse amplitude $I_0$ and should be inversely proportional to the pulse ratio $T/\tau$. 
This feature of the IP regime can be used to unambiguously detect one of the three parameters of the input pulse signal (its amplitude $I_0$, its duration $\tau$ or its repetition period $T$) using only the known (measured) value of $U_{\rm dc}$ if the other two parameters are known.
However, a potential limitation of this ``pulse recognition technique'' is that $U_{\rm dc}$ depends on the pulse ratio $T/\tau$. 
For input signals with equal pulse ratios $T/\tau$, but different pulse durations $\tau$ and repetition periods $T$, the value of $U_{\rm dc}$ should be the same.

\begin{figure*}
\includegraphics[width=1.0\textwidth]{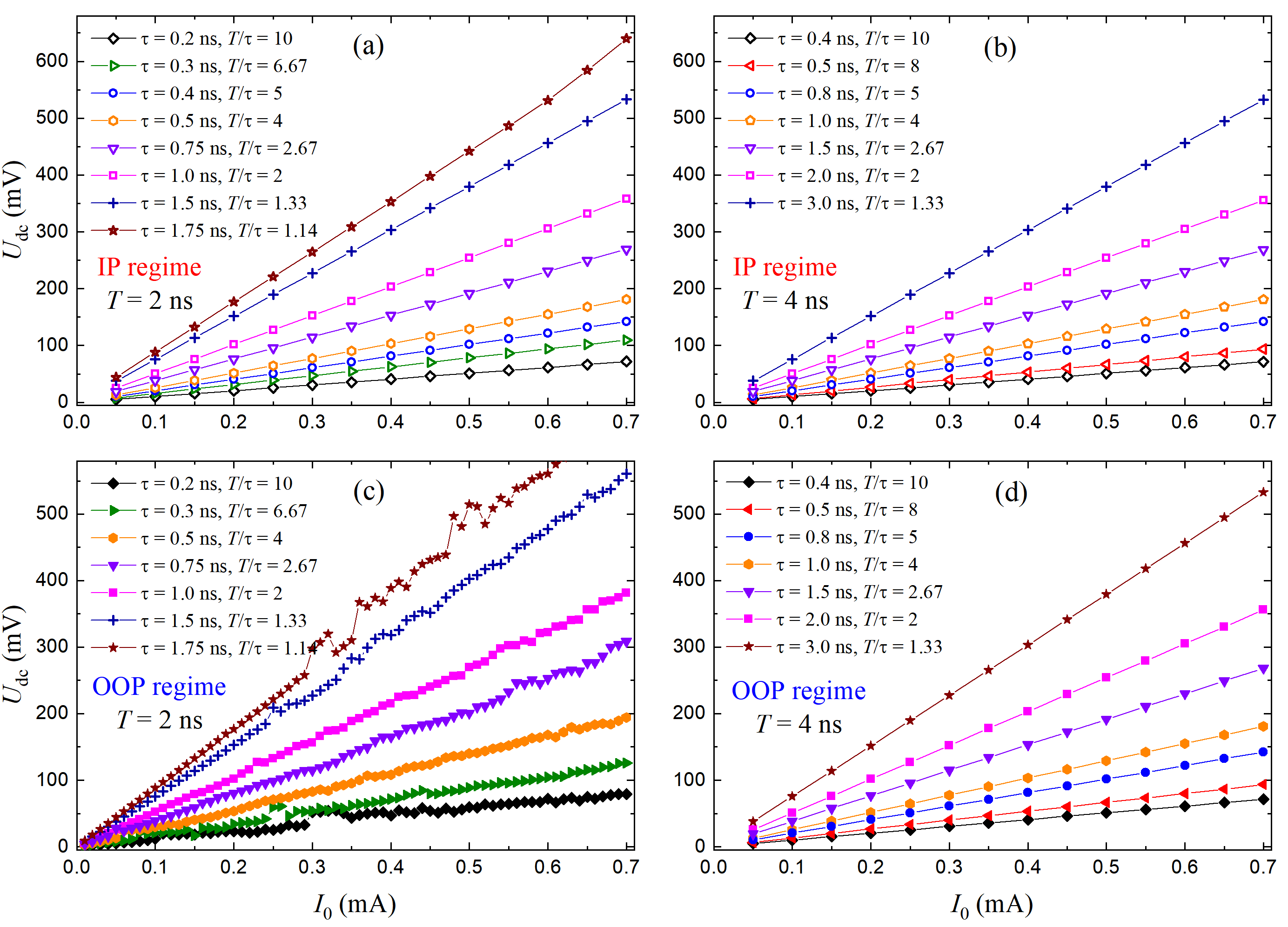}
\caption{
The detector output dc voltage $U_{\rm dc}$ as a function of the amplitude $I_0$ of input pulse current calculated numerically for different pulse durations $\tau$ and a pulse repetition period of $T = 2$~ns (a, c) and $T = 4$~ns (b, d). 
All calculations were performed for an STMD with typical parameters (see the main text for details), with IP magnetization dynamics (a, b) and OOP magnetization dynamics (c, d).
The curves shown in (a, b, d) demonstrate good agreement with the approximate linear dependence $U_{\rm dc} \sim I_0 (\tau/T)$, however, the curves in (c) contain voltage jumps and drops, which are typical for the nonlinear regime of device operation. 
To simplify the analysis, the curves with the same $T/\tau$ ratio are drawn in the same color.
}
\label{f:I0Udc}
\end{figure*}

As expected, the behavior of the time-dependent voltage $U(t)$ in the OOP regime is more complicated than in the IP regime.
Due to the large oscillations of $m_x(t)$ and $R(t)$, the voltage pulses have a non-flat top with significantly varying voltage values. 
However, they certainly have the same duration $\tau$ and period $T$ as the input signal (see the dashed blue line in Fig.~\ref{f:ImxRU}(e)). 
Despite the clearly visible voltage oscillations when $I(t) > 0$, the voltage behavior over time, $U(t)$, almost repeats itself every repetition period $T$ of the input signal. 
This quasi-periodicity of $U(t)$ allows one to obtain an approximated equation for $U_{\rm dc}$ similar to that for the IP regime.
According to Fig.~\ref{f:ImxRU}(c), it is natural to present $m_x(t)$ as a superposition of harmonic signals: $m_x(t) \approx m_1 \cos(\omega_1 t + \phi_1) + m_2 \cos(\omega_2 t + \phi_2) + \ldots + m_j \cos(\omega_j t + \phi_j) + \ldots$, where $m_j$, $\omega_j$ and $\phi_j$ are the amplitude, angular frequency and initial phase of the $j$th harmonic, $j \in [1, \infty)$.
Substituting this expression into (\ref{e:Udc}) yields $U_{\rm dc} = (I_0 R_\bot / T) \mathcal{I}$.
Here we took into account the repeatability of $m_x(t)$ and $U(t)$ oscillations, and introduce the integral $\mathcal{I} = \int_{t_0}^{t_0+\tau} dt/[1 + \eta^2 \{m_1 \cos(\omega_1 t + \phi_1) + m_2 \cos(\omega_2 t + \phi_2) + \ldots + m_j \cos(\omega_j t + \phi_j) + \ldots\}]$. 
We assume that the value of $\mathcal{I}$ is proportional to the pulse duration $\tau$. 
Then, writing the integral in the form $\mathcal{I} = \tau q$, we obtain the final equation for $U_{\rm dc}$ -- the equation (\ref{e:UdcApprox}), where $q = \mathcal{I}/\tau$ is a dimensionless quantity.
The equation (\ref{e:UdcApprox}) differs for the IP and OOP regimes by only a factor of $q$. 
When $q$ does not depend on $I_0$ and $\tau$, i.e., in the weakly nonlinear or linear regimes, the known (e.g., measured) value of $U_{\rm dc}$ can be used to unambiguously detect the input pulse parameters.
However, in the strongly nonlinear regime, when $q$ is a function of $I_0$ and/or $\tau$, the dependence of $U_{\rm dc}$ on $I_0$ and $\tau$ is nonlinear, which prevents the determination of the input pulse parameters from a known $U_{\rm dc}$.

To verify the results of our analysis, we numerically calculated the dependence of the output dc voltage $U_{\rm dc}$ on the input pulse amplitude $I_0$ for different pulse durations $\tau$ and pulse repetition periods $T$ as shown in Fig.~\ref{f:I0Udc}.
As can clearly be seen in Fig.~\ref{f:I0Udc}(a,b), the dependence of $U_{\rm dc}$ on $I_0$ is linear for any pulse duration $\tau$ in the IP regime. 
This agrees quite well with the theoretical dependence (\ref{e:UdcApprox}) for the IP regime, $U_{\rm dc} \approx I_0 R_0 (\tau/T)$, discussed earlier. 
Note that only one data point, at $I_0 = 0.7$~mA and $\tau = 1.75$~ns in Fig.~\ref{f:I0Udc}(a), does not precisely fit the linear dependence (\ref{e:UdcApprox}).
Additionally, comparing the values of $U_{\rm dc}$ calculated for the same $I_0$ and $T/\tau$, but different values of $\tau$ and $T$, one can conclude that the $U_{\rm dc}$ value is precisely proportional to $\tau/T$ (the curves with the same $T/\tau$ are shown in the same colors in Fig.~\ref{f:I0Udc}(a, b)).

In the OOP regime, the dependence $U_{\rm dc}(I_0)$ can be considered as precisely linear for a large pulse repetition period of $T = 4$~ns (Fig.~\ref{f:I0Udc}(d)). 
This case corresponds to the previously obtained equation (\ref{e:UdcApprox}) with $q \approx 1.02$. 
However, for input current pulses with a smaller repetition period of $T = 2$~ns, a more complicated, nonlinear dependence $U_{\rm dc}(I_0)$ is observed (Fig.~\ref{f:I0Udc}(c)).
At small values of $I_0$, the dependence $U_{\rm dc}(I_0)$ appears quite linear. 
However, above a threshold, when $I_0 \ge I_{\rm th}$, significant changes occur in the dependence of $U_{\rm dc}(I_0)$, including voltage jumps or drops.
For instance, this threshold is clearly seen at $I_{\rm th} \approx 0.29$~mA for the top curve in Fig.~\ref{f:I0Udc}(c), which was calculated for $\tau = 1.75$~ns and $T = 2$~ns. 
The magnitude of these voltage jumps or drops depends on the voltage value (it can be higher for higher voltage values). 
Our preliminary analysis indicates that these voltage deviations from linear dependence can be explained by the specific features of magnetization dynamics excited by the input current, however, this requires additional study.

The linearity of the dependence $U_{\rm dc}$ on $I_0 \tau/T$ observed in the IP regime (Fig.~\ref{f:I0Udc}(a, b)) and partially in the OOP regime (Fig.~\ref{f:I0Udc}(d) and when $I_0 \le I_{\rm th}$ in Fig.~\ref{f:I0Udc}(c)), demonstrates the possibility of unambiguous detection of the parameters of positive rectangular pulse signals in an STMD. 
Using a single $U_{\rm dc}$ value, one can determine one of the three input signal parameters -- the pulse amplitude $I_0$, its duration $\tau$, and repetition period $T$ -- if other two pulse parameters are known. 
However, the efficiency of this technique is severely limited by the linear relationship between $U_{\rm dc}$, $I_0$, and $\tau/T$. 
Therefore, the IP regime of STMD operation is preferable for pulse detection.

In conclusion, we have demonstrated that an STMD with an applied input positive rectangular pulse current can operate in two distinct regimes: linear and nonlinear. 
In the linear regime, which is easily observed in an STMD with IP magnetization dynamics, the detector's output dc voltage $U_{\rm dc}$ is proportional to the pulse amplitude $I_0$ and its duration $\tau$ and inversely proportional to the pulse repetition period $T$, which can be used for the unambiguous determination of the input pulse signal parameters. 
In the nonlinear regime, which is revealed in an STMD with OOP magnetization dynamics and is excited by input pulses with a sufficiently large amplitude $I_0 \ge I_{\rm th}$, the generated dc voltage undergoes voltage jumps and drops, which prevents the unambiguous determination of the input signal parameters.
The developed formalism can be useful for the development and optimization of spintronic devices capable of detecting and processing non-harmonic (e.g., digital) microwave signals.

This work was supported by grant No.~2025.07/0237 from the National Research Foundation of 
Ukraine. 
The authors would like to thank all the brave defenders of Ukraine who made the finalization of this publication possible.


\end{document}